\documentstyle[eqsecnum,prd,aps,preprint,epsf]{revtex}

\draft
\begin{document}

\def\beqa{\begin{eqnarray}}
\def\eeqa{\end{eqnarray}}
\def\bI{\hbox{$\,I\!\!\!\!-$}}
\def\a{\alpha}
\def\b{\beta}
\def\p{\partial}
\def\e{\epsilon}
\def\ve{\varepsilon}
\def\r{\rho}
\def\O{\Omega}
\def\t{\tilde}
\def\ra{\rightarrow}

\thispagestyle{empty}
{\baselineskip0pt
\leftline{\large\baselineskip16pt\sl\vbox to0pt{\hbox{\it Department of Physics}
               \hbox{\it Kyoto University}\vss}}
\rightline{\baselineskip16pt\rm\vbox to20pt{\hbox{KUNS 1425}
           \hbox{OU-TAP 55}
               \hbox{\today}
\vss}}%
}
\vskip1cm
\begin{center}{\large \bf
Solving the Darwin problem in the first post-Newtonian approximation of 
general relativity}
\end{center}

\vskip 0.4cm

\begin{center}
Keisuke Taniguchi 
\footnote{e-mail: taniguci@tap.scphys.kyoto-u.ac.jp}~
and ~
Masaru Shibata 
\footnote{e-mail: shibata@vega.ess.sci.osaka-u.ac.jp}
\end{center} 

\begin{center}
$^*${\em Department of Physics,~Kyoto University,~Kyoto 606-01,~Japan} \\
$^\dagger$
{\em Department of Earth and Space Science,~Osaka University,~Toyonaka 
560,~Japan}
\end{center}

\begin{abstract}
We analytically calculate the equilibrium sequence of the corotating 
binary stars of incompressible fluid 
in the first post-Newtonian(PN) approximation of general relativity. 
By calculating the total energy
and total angular momentum of the system as a function of the orbital 
separation, we investigate the 
innermost stable circular orbit for corotating binary(we call it ISCCO).
It is found 
that by the first PN effect, the orbital 
separation of the binary at the ISCCO becomes small with increase of 
the compactness of each star, and as a result, 
the orbital angular velocity at the ISCCO increases. These 
behaviors agree with previous numerical works. 
\end{abstract}

\newcommand{\beq}{\begin{equation}}
\newcommand{\eeq}{\end{equation}}
\newcommand{\beqn}{\begin{eqnarray}}
\newcommand{\eeqn}{\end{eqnarray}}
\newcommand{\pa}{\partial}

\section{Introduction}

The laser interferometers such as LIGO\cite{LIGO}, VIRGO\cite{VIRGO},
GEO600\cite{GEO} and TAMA300\cite{TAMA} are currently being constructed
and the detection of gravitational waves is expected in this
decade. One of the most important astrophysical sources of
gravitational waves for these detectors is the coalescing binary neutron
stars(BNS's) because the gravitational waves emitted in the inspiral phase
(the so-called last three minutes)\cite{LTM} have frequencies in the sensitive
region of these detectors, i.e., from 10Hz to 1000Hz. 
We will be able to know each 
mass, spin, and so on of the BNS's if we could obtain 
an accurate theoretical template for data analysis\cite{CF}. 
Hence, much theoretical works have been done to complete 
it\cite{BDIWW}\cite{kidder}\cite{Blanchet}. 

When the orbital separation of the BNS's becomes 
a few times of the neutron star(NS) radius as a result of the radiation 
reaction of gravitational wave emission, the hydrodynamical effect 
becomes important. In such a phase, the wave form of gravitational waves 
is expected to be sensitive to the NS structure, especially the 
relation between the radius and mass of the NS. Therefore, if gravitational
waves from such a phase are detected, 
we may constrain the equation of state(EOS) of the NS\cite{lindblom}
\cite{joan}\cite{LRSLRS}.
In particular, the important quantity is the location of 
the innermost stable circular orbit(ISCO), which will have an 
information on the EOS of the NS. 

There have been many analyses for the ISCO using some approximations of 
general relativity\cite{KWW}\cite{WS}\cite{LW}\cite{TN}, but in order 
to determine the precise location of the ISCO, we need fully 
general relativistic(GR) numerical simulation, 
which is a very difficult method. 
Recently, Wilson and his collaborators\cite{wilson} have performed 
numerical simulation for obtaining the equilibrium sequence of 
the BNS and its ISCO solving semi-relativistic equations. 
The results they have obtained are very interesting, but the 
accuracy of their results is still in question because they do not 
show any calibration of their numerical code using test problems. 
Furthermore, they do not seem to perform any important analysis to their
numerical results. 
When we carry out a large numerical simulation, it is required to perform a 
detailed analysis after the computing in order 
to explain the numerical results. 
The analysis is desired to be done by comparing the analytical or 
semi-analytical estimates. 
If the numerical results qualitatively 
agree with such an analytical calculation, 
we can firmly believe the numerical results and also can understand 
details. Hence, with a large scale simulation, 
it is favorable to prepare some analytical 
models in order to understand the essence included in the numerical results. 

For that purpose, we here analytically solve the Darwin problem 
in the first PN approximation. The Darwin
problem is concerned with the equilibrium and the stability of a
homogeneous fluid star rotating around another one taking into 
account the mutual tidal interactions\cite{Ch69}. We assume that each 
NS in the binary system is composed of the incompressible and 
homogeneous fluid. 
The reward is that all the calculations can be done analytically. 
This means that we can obtain a strict solution of a BNS 
including the GR effect without any large 
supercomputing. Our results will be very helpful for understanding how the
finite-size effects as well as the GR one of the NS's 
influence to the location and the orbital angular velocity at 
the ISCO, and for the analysis of large scale simulation. 

This paper is organized as follows. In section II, we show the basic equations 
to solve the PN Darwin problem. By using the 
first tensor virial(TV) equations, we 
derive the angular velocity of corotating binary systems in section III. 
In section IV, we show the equations of the total energy and the total 
angular momentum for corotating binary systems. In section V, we calculate 
the equilibrium sequences and determine the location of the energy and 
angular momentum minimums(we call it the innermost stable corotating 
circular orbit(ISCCO) to distinguish it from the innermost stable
circular orbit(ISCO)\footnote{This point is the secular instability
  limit, and not the dynamical instability limit(i.e., ISCO). The ISCO
  will locate inside the ISCCO\cite{LRS}.}). Section VI is devoted to
summary.

Throughout this paper, we use the unit of $G=1$, and $c$ denotes 
the light velocity. 
Latin indices $i,j,k,\cdots$ take 1 to 3, and $\delta_{ij}$ 
denotes the Kronecker's delta. We use $I_{ij}$ and $\bI_{ij}$ 
as the quadrupole moment and its trace free part, 
\beq
\bI_{ij}=I_{ij}-{1 \over 3}\delta_{ij}\sum_{k=1}^3 I_{kk}, 
\eeq
of each star of binary.

\section{Formulation}

Non-axisymmetric equilibrium configurations 
of uniformly rotating incompressible fluid in the
first PN approximation are obtained 
by solving the integrated form of the Euler equation and the Poisson 
equations for gravitational potentials consistently. 
The integrated form of the Euler equation 
was derived by Chandrasekhar\cite{PNeom} and it can be written 
as\cite{shibapn}\cite{asada} 
\beqa
  & &\frac{P}{\r}-\frac{1}{2c^2} \left( \frac{P}{\r} \right)^2 \nonumber \\
  &=& U-\frac{X_0}{c^2} + \left\{ \frac{\varpi^2}{2} + \frac{1}{c^2}
  \left( 2\varpi^2 U-X_{\O} + \hat{\beta}_{\varphi} \right) \right\} \O^2
  + \frac{\varpi^4}{4 c^2} \O^4 + {\rm const.}, \label{ftve}
\eeqa
where
\beqa
  \varpi^2 &=& \left(x_1 + \frac{R}{2} \right)^2 +x_2^2,
\eeqa
and $U$, $X_0$, $X_{\O}$, and $\hat \beta_{\varphi}$ are the 
gravitational potentials. 
In this paper, we consider the equilibrium sequences of BNS's 
of equal masses($M_1=M_2=M$) 
whose coordinate separation is $R$\footnote{The coordinate 
condition in this paper is the standard PN one\cite{chand}.}. We assume that
the center of mass of a star (star 1) locates at the origin and the
other one (star 2) locates at $(x_1,~x_2,~x_3)=(-R,~0,~0)$(see fig.1). 
Due to the symmetry, we only pay attention to the equilibrium 
configuration of star 1 in the following. 

The main purpose of this paper is to calculate the PN correction of the 
angular velocity, the energy and the angular momentum for the 
Darwin problem. In the Newtonian order, the angular velocity($\Omega_{\rm N}$)
 becomes\cite{LRS}(see below for derivation)
\beq
\Omega^2_{\rm N}={2 M \over R^3}+{18\bI_{11} \over R^5}. \label{omen}
\eeq
Thus, in the PN approximation, we can expect that the 
following types of quantities will be the main terms in the PN order:
\beq
\sim {M \over R^3}\times {M \over a_0c^2},~~~~~
\sim {M \over R^3}\times {M \over R  c^2},~~~~~
\sim {Ma_0^2 \over R^5}\times {M \over a_0c^2},~~~{\rm and}~~~
\sim {Ma_0^2 \over R^5}\times {M \over Rc^2},
\eeq
where $a_0$ is a typical radius of the star, and we use the relation 
$\bI_{11} \sim Ma_0^2$. We will derive four types of 
terms shown above and the correction of the energy and the angular momentum 
by them below. 

Since we consider the incompressible fluid, the gravitational potentials 
inside each star 
are expressed as the polynomial form of the coordinates $x_i$. 
For the purpose of obtaining the PN corrections shown above, 
we need to take into account the coefficients of the terms such as
$x_1^{m_1}x_2^{m_2}x_3^{m_3}$ in $X_0$, $\O^2 X_{\O}$, $\O^2
\hat{\beta}_{\varphi}$, and so on, where $0 \le m_1, m_2, m_3 \le 5$ and 
$0 \le m_1+m_2+m_3 (\equiv m_t) \le 5$, up to $O(R^{-5})$ for the case
$m_t$ is odd and up to $O(R^{-3})$ for the case $m_t$ is even.
In the following subsections, we solve the 
Poisson equations for the gravitational potentials to derive such terms. 

\subsection{Newtonian Quantities}

Each gravitational potential 
is composed of two parts; one is the contribution from star 1 and 
the other is from star 2. In the following, we denote the former part 
as $\phi^{1 \rightarrow 1}$, and the latter one as 
$\phi^{2 \rightarrow 1}$, where $\phi$ denotes one of the potentials. 
We also define $\phi^{1 \rightarrow 2}$ and 
$\phi^{2 \rightarrow 2}$ as the contribution from star 1 to 2 and 
star 2 itself, respectively. 

Following previous authors\cite{Ch69}\cite{LRS}, the configuration of 
each star of binary in the Newtonian order is assumed to be 
an ellipsoidal figure of its axial length $a_1$, $a_2$ and $a_3$. 
In this case, 
the solution of the Poisson equation for the Newtonian potential 
\beqa
  \Delta U= -4 \pi \r,
\eeqa
is written as $U=U^{1 \ra 1} +U^{2 \ra 1}$, where 
\beqa
  U^{1 \ra 1} &=& \pi \r \left( A_0 - \sum_l A_l x_l^2 \right), \\
  U^{2 \ra 1} &=& \frac{M}{R} \left\{ 1-\frac{x_1}{R} +\frac{2x_1^2
      -x_2^2 -x_3^2}{2R^2} + \frac{-2x_1^3 +3x_1(x_2^2 +x_3^2)}{2R^3}
    \right. \nonumber \\
  & &\hspace{40pt} \left.+
    \frac{8x_1^4 +3x_2^4 +3x_3^4 -24 x_1^2 (x_2^2 +x_3^2) +6x_2^2
      x_3^2}{8R^4} \right\} \nonumber \\
  & &+ \frac{3 \bI_{11}}{2R^3} \left( 1-\frac{3x_1}{R} +\frac{12x_1^2
      -5x_2^2 -5x_3^2}{2R^2} \right) +\frac{3}{2R^5} \left( \bI_{22}
    x_2^2 +\bI_{33} x_3^2 \right), \label{Uout}
\eeqa
and 
\beqa
  I_{ijkl \cdots} &=& \int d^3 x \r x_i x_j x_k x_l \cdots. 
\eeqa
$A_{ij \cdots}$ are index symbols introduced by
Chandrasekhar\cite{Ch69}, and $A_0 =\sum_l A_l a_l^2$ 
is calculated from\cite{Ch69}
\beqa
  A_0 &=& a_1 a_2 a_3 \int_0^{\infty}
  \frac{du}{\sqrt{(a_1^2+u)(a_2^2+u)(a_3^2+u)}} \\
  &=& a_1^2 \a_2 \a_3 \int_0^{\infty}
  \frac{dt}{\sqrt{(1+t)(\a_2^2+t)(\a_3^2+t)}} \equiv a_1^2 \tilde A_0,
\eeqa
where $\a_2 =a_2/a_1$ and $\a_3 =a_3/a_1$. Note that $U^{2 \ra 2}$ and
$U^{1 \ra 2}$ are obtained by changing $x_1$ into $-(x_1+R)$ in $U^{1
  \ra 1}$ and $U^{2 \ra 1}$, respectively.

In the Newtonian order, the pressure is written as
\beqa
  P=P_0 \left( 1- \sum_l \frac{x_l^2}{a_l^2} \right). 
\eeqa
$P_0$ is calculated from the scalar virial relation as 
\beqa
  P_0 = \frac{\r}{3} \left[ \pi \r A_0 -\frac{\O_{{\rm N}}^2}{2}
  \left( a_1^2 +a_2^2 \right) -\frac{M}{2R^3} \left( 2a_1^2 -a_2^2
    -a_3^2 \right) \right] +O(R^{-5}), \label{P_0}
\eeqa
and $\alpha_2$ and $\alpha_3$ are determined from\cite{LRS} 
\beqa
  -\frac{P_0}{\r a_1^2} &=& -\pi \r A_1+\frac{\O_{\rm
      N}^2}{2}+\frac{M}{R^3}, \label{P_1} \\
  -\frac{P_0}{\r a_2^2} &=& -\pi \r A_2+\frac{\O_{\rm
      N}^2}{2}-\frac{M}{2R^3}, \label{P_2} \\
  -\frac{P_0}{\r a_3^2} &=& -\pi \r A_3 -\frac{M}{2R^3}. \label{P_3}
\eeqa

\subsection{$X_0$}

As in the case of $U$, the PN potentials 
are divided into two parts as $X_0=X_0^{1\ra 1} +X_0^{2 \ra 1}$, 
and we consider them separately. 

\begin{itemize}
\item Contribution from star 1:

$X_0^{1 \ra 1}$ is derived from the Poisson equation\cite{shibapn}
\beqa
  \Delta X_0^{1 \ra 1} = 4\pi \r \left[ 2 \pi \r \left(A_0 -\sum_l A_l x_l^2
    \right) +\frac{3P_0}{\r} \left( 1- \sum_l \frac{x_l^2}{a_l^2} \right) 
    +2 U^{2 \ra 1} \right],
\eeqa
and the solution becomes
\beqn
  X_0^{1 \ra 1} &=& -\a_0 U^{1 \ra 1} +\a_1 D_1 +\sum_l
  \eta_l D_{ll} \nonumber \\
&&  -\frac{M}{R^3} \Bigl(2D_{11} -D_{22} -D_{33}\Bigr)
    -\frac{M}{R^4} \Bigl(-2D_{111} +3D_{122} +3D_{133}\Bigr),
\eeqn
where
\beqa
  \a_0 &=& 2 \pi \r A_0 + \frac{3P_0}{\r} +\frac{2M}{R} +\frac{3
    \bI_{11}}{R^3}, \\
  \a_1 &=& \frac{2M}{R^2} +\frac{9 \bI_{11}}{R^4}, \\
  \eta_l &=& 2\pi \r A_l +\frac{3P_0}{\r a_l^2}.
\eeqa
$D_i$, $D_{ii}$, and $D_{1ii}$ are the solutions of equations
\beqa
  \Delta D_i &=& -4 \pi \r x_i, \\
  \Delta D_{ii} &=& -4 \pi \r x_i^2, \\
  \Delta D_{1ii} &=& -4 \pi \r x_1 x_i^2,
\eeqa
and the solutions at star 1 are\cite{Ch69}
\beqa
  D_i &=& \pi \r a_i^2 \left( A_i -\sum_l A_{il} x_l^2 \right) x_i , \\
  D_{ii} &=& \pi \r \left[ a_i^4 \left( A_{ii} -\sum_l A_{iil} x_l^2
  \right) x_i^2 \right. \nonumber \\
  & &\hspace{40pt} \left. + \frac{1}{4} a_i^2 \left( B_i -2\sum_l B_{il}
  x_l^2 + \sum_l \sum_m B_{ilm} x_l^2 x_m^2 \right) \right], \\
  D_{111} &=& \pi \r \left[ a_1^6 \left( A_{111} -\sum_l A_{111l} x_l^2
  \right) x_1^3 \right. \nonumber \\
  & &\hspace{40pt} \left.+ \frac{3}{4} a_1^4 \left( B_{11} -2 \sum_l
  B_{11l} x_l^2 +
  \sum_l \sum_m B_{11lm} x_l^2 x_m^2 \right) x_1 \right], \\
  D_{1ii} &=& \pi \r \left[ a_1^2 a_i^4 \left( A_{1ii} -\sum_l A_{1iil}
  x_l^2 \right) x_1 x_i^2 \right. \nonumber \\
  & &\hspace{40pt} \left. +\frac{1}{4} a_1^2 a_i^2 \left( B_{1i} -2\sum_l
  B_{1il} x_l^2 + \sum_l \sum_m B_{1ilm} x_l^2 x_m^2 \right) x_1\right], 
\eeqa
where $B_{ijk \cdots}$ are index symbols defined by
Chandrasekhar\cite{Ch69}.

\item Contribution from star 2:

The equation for $X_0^{2 \ra 1}$ is
\beqa
  \Delta X_0^{2 \ra 1} = 4\pi \r \left[ 2 \pi \r \left(A_0 -\sum_l A_l y_l^2
    \right) +\frac{3P_0}{\r} \left( 1- \sum_l \frac{y_l^2}{a_l^2} \right) 
    +2 U^{1 \ra 2} \right],
\eeqa
where $y_1 =-(x_1+R)$, $y_2=x_2$ and $y_3=x_3$. The solution is written as
\beq
  X_0^{2 \ra 1} = -\a_0 U^{2 \ra 1} -\a_1 D_1^{2 \ra 1}
      +\sum_l \eta_l D_{ll}^{2 \ra 1} 
  -\frac{M}{R^3} \Bigl(2D_{11}^{2 \ra 1} - D_{22}^{2 \ra 1}
  -D_{33}^{2 \ra 1}\Bigr), \label{X_0^(2)}
\eeq
where $D_{ij \cdots}^{2 \ra 1}$ are calculated from the same equations
as the case of $D_{ij \cdots}$\cite{Ch69}, i.e.,
\beqa
  \Delta D_{ij \cdots}^{2 \ra 1} =-4 \pi \r y_i y_j \cdots.
\eeqa
The solutions of $D_{ij \cdots}^{2 \ra 1}$ are 
\beqa
  D_1^{2 \ra 1} &=& \frac{I_{11}}{R^2} \left(1-\frac{2x_1}{R}
    +\frac{6x_1^2-3x_2^2-3x_3^2}{2R^2} + O(R^{-3}) \right), \\
  D_2^{2 \ra 1} &=& \frac{I_{22}}{R^2} \left( \frac{x_2}{R}
    -\frac{3x_1x_2}{R^2} + O(R^{-3}) \right), \\
  D_{ii}^{2 \ra 1} &=& \frac{I_{ii}}{R} \left( 1-\frac{x_1}{R}
    +\frac{2x_1^2 -x_2^2 -x_3^2}{2R^2}
    +\frac{-2x_1^3+3x_1(x_2^2+x_3^2)}{2R^3} + O(R^{-4}) \right)
    \nonumber \\
   & &+\frac{3 \bI_{ii11}}{2R^3} 
   \left( 1-\frac{3x_1}{R} + O(R^{-2}) \right),
\eeqa
where
\beqa
  \bI_{ii11} = I_{ii11} -\frac{1}{3} \sum_lI_{iill}.
\eeqa

\end{itemize}

\subsection{$X_{\O}$}

\begin{itemize}

\item Contribution from star 1:

The equation for $X_{\O}^{1 \ra 1}$ is 
\beqa
  \Delta X_{\O}^{1 \ra 1} =8 \pi \r \left( x_1^2 +x_2^2 +Rx_1
    +\frac{R^2}{4} \right).
\eeqa
Then the solution is
\beqa
  X_{\O}^{1 \ra 1} =-2 \left( D_{11} +D_{22} +RD_1 +\frac{R^2}{4} U^{1
      \ra 1} \right).
\eeqa

\item Contribution from star 2:

The equation $X_{\O}^{2 \ra 1}$ may be written as 
\beqa
  \Delta X_{\O}^{2 \ra 1} =8 \pi \r \left( y_1^2 +y_2^2 -Ry_1
    +\frac{R^2}{4} \right).
\eeqa
Then, using $D_{ij \cdots}^{2 \ra 1}$, the solution is easily 
derived as
\beqa
  X_{\O}^{2 \ra 1} &=&-2 \left( D_{11}^{2 \ra 1} +D_{22}^{2 \ra 1}
    -RD_1^{2 \ra 1} +\frac{R^2}{4} U^{2 \ra 1} \right), \\
  &=& -\frac{R^2}{2} U^{2 \ra 1} -\frac{2I_{11}}{R^2} x_1 -
  \frac{2I_{22}}{R} \left( 1- \frac{x_1}{R} \right).
\eeqa

\end{itemize}

\subsection{$\hat{\beta}_{\varphi}$}

The definition of $\hat{\beta}_{\varphi}$ is\cite{shibapn}
\beqa
  \hat{\beta}_{\varphi} &=& -\frac{7}{2} \left( x_1 P_1 +x_2 P_2
    +\frac{R}{2} P_1 \right) \nonumber \\
  & &-\frac{1}{2} \left[ \left( x_1 +\frac{R}{2} \right)^2 P_{2,2}
    +x_2^2 P_{1,1} -\left( x_1+
    \frac{R}{2} \right) x_2 (P_{1,2} +P_{2,1} ) \right], \label{betaphi}
\eeqa
where $P_1$ and $P_2$ satisfy, 
\beqa
  \Delta P_1 &=&-4 \pi \r \left( x_1 +\frac{R}{2} \right),
  \\
  \Delta P_2 &=&-4 \pi \r x_2.
\eeqa
$P_i$ is also written as $P_i^{1 \ra 1} +P_i^{2 \ra 1}$, where
\beqa
  P_1^{1 \ra 1} &=& D_1 +\frac{R}{2} U^{1 \ra 1}, \\
  P_2^{1 \ra 1} &=& D_2,
\eeqa
and
\beqa
  P_1^{2 \ra 1} &=& D_1^{2 \ra 1} -\frac{R}{2} U^{2 \ra 1}, \label{P_1^2}\\
  P_2^{2 \ra 1} &=& D_2^{2 \ra 1}. \label{P_2^2}
\eeqa

\subsection{The Collection} 

Substituting expressions for the gravitational potentials 
derived in subsections {\bf A}$-${\bf D} into Eq.(\ref{ftve}), the 
integrated form of the Euler equation is written as
\beqa
  \frac{P}{\r} &=& U + \delta U +\frac{1}{2} \varpi^2 \O^2 \nonumber \\
  & &+\frac{1}{c^2} \left[ \gamma_0 +\sum_l \gamma_l x_l^2 +\sum_{l \ge m}
    \gamma_{lm} x_l^2 x_m^2 +x_1 \left( \beta_0 + \sum_l \beta_l x_l^2 + 
      \sum_{l \ge m} \beta_{lm} x_l^2 x_m^2 \right) \right] \nonumber \\
  & &+{\rm const.},
  \label{ieom}
\eeqa
where $\delta U$ is the PN correction of $U$ which we will mention in the
next section. In the following, we do not need $\gamma_{0}$, $\gamma_{i}$, 
and $\gamma_{ij}$(see below), 
but need $\beta_0$, $\beta_{i}$ and $\beta_{ij}$, 
which are 
\beqa
  \beta_0 &=& -\frac{M \pi \r}{R^2} \left( \frac{6P_0}{5\pi \r^2}
    +\frac{11A_0}{10} +\frac{3}{2} a_1^2 A_1 +a_2^2 A_2 \right)
  +\frac{9M^2}{4R^3} \nonumber \\
  & &+\frac{1}{R^4} \Biggl[ \frac{9}{2} \pi \r \bI_{11}
    (-a_1^2 A_1 -2a_2^2 A_2 +A_0) +\frac{9}{2} \sum_l \left( 2\pi \r
      A_l + \frac{3P_0}{\r a_l^2} \right) \bI_{ll11} \nonumber \\
  & &\hspace{40pt} +\frac{3}{4} \pi \r
      Ma_1^2 (-2a_1^2 B_{11} +a_2^2 B_{12} +a_3^2 B_{13}) -\frac{9}{2}
      \pi \r \bI_{11} \left( 2A_0 +\frac{3P_0}{\pi \r^2} \right) \Biggr]
  \nonumber \\
  & & +\frac{M}{8R^5} (118I_{11} -93I_{22} -59I_{33}), \\
  \beta_1 &=& \frac{M\pi \r}{R^2} \left( \frac{3}{2} a_1^2 A_{11} +a_2^2
    A_{12} -\frac{1}{2} A_1 \right) \nonumber \\
  & &+\frac{1}{R^4} \Biggl[ \frac{9}{2} \pi \r
    \bI_{11} (a_1^2 A_{11} +2a_2^2 A_{12} -A_1) +\sum_l \eta_l I_{ll}
    -\left( 2\pi \r A_0 +\frac{3P_0}{\r} \right) M \nonumber \\
  & &\hspace{40pt} +\frac{1}{2} \pi \r M 
    a_1^2 (-4a_1^4 A_{111} +6a_1^2 B_{111} -3a_2^2 B_{112} -3a_3^2
    B_{113}) \Biggr] -\frac{M^2}{4R^5}, \\
  \beta_2 &=& \frac{M\pi \r}{R^2} \left( \frac{1}{2} a_1^2 A_{12} -a_2^2 
    A_{12} +3a_2^2 A_{22} +A_1 -\frac{3}{2} A_2 \right)
    \nonumber \\
  & &+\frac{1}{R^4} \Biggl[ \frac{9}{2} \pi \r \bI_{11} (-a_1^2 A_{12}
    -2a_2^2 A_{12} +6a_2^2 A_{22} -3A_2 +2A_1) \nonumber \\
  & &\hspace{40pt} -\frac{3}{2} \sum_l \eta_l 
    I_{ll} +\frac{3}{2} \left( 2\pi \r A_0 +\frac{3P_0}{\r} \right) M
    \nonumber \\
  & &\hspace{40pt} -\frac{3}{2} \pi \r M a_1^2 (-2 a_1^2
    B_{112}-2a_2^4 A_{122}+a_2^2 B_{122}
    +a_3^2 B_{123}) \Biggr] +\frac{29M^2}{8R^5}, \\
  \beta_3 &=& \frac{M\pi \r}{R^2} \left( \frac{3}{2} a_1^2 A_{13}
    +a_2^2 A_{23} -\frac{1}{2} A_3 \right) \nonumber \\
  & &+\frac{1}{R^4} \Biggl[ \frac{9}{2} \pi \r \bI_{11} (a_1^2 A_{13}
    +2a_2^2 A_{23} -A_3)
    -\frac{3}{2} \sum_l \eta_l I_{ll} +\frac{3}{2} \left( 2\pi \r A_0
      +\frac{3P_0}{\r} \right) M \nonumber \\
  & &\hspace{40pt} -\frac{3}{2} \pi \r M a_1^2 (-2a_1^2
    B_{113} -2a_3^4 A_{133} +a_2^2 B_{123} +a_3^2 B_{133}) \Biggr]
    +\frac{39M^2}{8R^5}, \\
  \beta_{11} &=& \frac{M\pi \r}{R^4} \left[ 2a_1^6 A_{1111} +\frac{3}{4} 
    a_1^2 (-2a_1^2 B_{1111} +a_2^2 B_{1112} +a_3^2 B_{1113}) \right], \\
  \beta_{22} &=& \frac{M\pi \r}{R^4} \left[ -3a_1^2 a_2^4 A_{1222}
    +\frac{3}{4} a_1^2 (-2a_1^2 B_{1122} +a_2^2 B_{1222} +a_3^2
    B_{1223}) \right], \\
  \beta_{33} &=& \frac{M\pi \r}{R^4} \left[ -3a_1^2 a_3^4 A_{1333}
    +\frac{3}{4} a_1^2 (-2a_1^2 B_{1133} +a_2^2 B_{1233} +a_3^2
    B_{1333}) \right], \\
  \beta_{12} &=& \frac{M\pi \r}{R^4} \left[ 2a_1^6 A_{1112} -3a_1^2
    a_2^4 A_{1122} +\frac{3}{2} a_1^2 (-2a_1^2 B_{1112} +a_2^2 B_{1122}
    +a_3^2 B_{1123}) \right], \\
  \beta_{13} &=& \frac{M\pi \r}{R^4} \left[ 2a_1^6 A_{1113} -3a_1^2
    a_3^4 A_{1133} +\frac{3}{2} a_1^2 (-2a_1^2 B_{1113} +a_2^2 B_{1123}
    +a_3^2 B_{1133}) \right], \\
  \beta_{23} &=& \frac{M\pi \r}{R^4} \left[ -3a_1^2 (a_2^4 A_{1223}
    +a_3^4 A_{1233}) +\frac{3}{2} a_1^2 (-2a_1^2 B_{1123}+a_2^2 B_{1223}
    +a_3^2 B_{1233}) \right].
\eeqa

\section{The Post-Newtonian angular velocity}

In this section, the orbital angular velocity in the PN 
order is derived using the first TV equation. 
The first TV relation is derived from 
\beqa
  \int d^3 x \frac{\p P}{\p x_1} =0.\label{tvtvtv}
\eeqa
Substituting Eq.(\ref{ieom}) to Eq.(\ref{tvtvtv}), we have
Eq.(\ref{omen}) for the Newtonian order. 
In the PN order, the explicit form becomes 
\beqa
  0 &=& \frac{R}{2} M \delta \O^2 +\int d^3 x \r \xi_1 \O_{{\rm N}}^2 +M
  \beta_0 +3\beta_1 I_{11} +\beta_2 I_{22} +\beta_3 I_{33} +5\beta_{11}
  I_{1111} \nonumber \\
  & &+3\beta_{12} I_{1122} +3\beta_{13} I_{1133} +\beta_{22} I_{2222}
  +\beta_{33} I_{3333} +\beta_{23} I_{2233} +\delta \int d^3 x \r
  \frac{\p U}{\p x_1}, \label{PNftve}
\eeqa
where $\xi_1$ is the $x_1$ component of the Lagrangian displacement
$\xi_i$, and $\delta \O^2$ denotes the PN correction of the orbital
angular velocity.

\subsection{The Post-Newtonian Deformation}

The density profile of an incompressible, homogeneous sphere in the PN
order is the same that that in the Newtonian order. However, the density
profile of an ellipsoid in the PN order is different from that in the
Newtonian order.
The Chandrasekhar's method to calculate
the PN effects on the uniformly rotating isolated
bodies is as follows\cite{PNeom}. First, he constructs the 
ellipsoidal figures in the Newtonian order. Then, he regards the PN effect 
as a small perturbation to the 
Newtonian configuration, and calculates the deformation from the Newtonian 
ellipsoid using the Lagrangian displacement. Solving the equations for the 
Lagrangian displacement and calculating the correction to $U$ by the 
deformation, he has obtained the equilibrium configuration in the PN order.
In this paper, we follow his method. 

Although the deformation from the ellipsoidal figure will really occur
by the PN effect, we will show in this subsection that it is a small 
effect compared with leading terms shown in Eq.(2.4). 
Hence, for readers who are not interested in such a verification, we 
recommend to skip to the next subsection. 

In order to express Eq.(\ref{ieom}) using only the coordinates and the
index symbols, we have to calculate $\delta U$ which is defined as 
\beqa
  \delta U =-\sum_i \frac{\p}{\p x_i} \int d^3 x' \frac{\r \xi_i}{|{\bf
      x} -{\bf x}'|}.
\eeqa
Here, $\xi_i$ denotes the Lagrangian displacement of the fluid element
induced by the PN correction, and is used to guarantee that the pressure 
of the deformed Darwin ellipsoid at the boundary becomes zero. This
means that it is determined from $\gamma_0$, $\gamma_i$, $\gamma_{ij}$,
$\beta_0$, $\beta_i$ and $\beta_{ij}$ in Eq.(\ref{ieom}) in general. 

In choosing $\xi_i$, we require it to be divergent free because $\delta
\r =-\sum_i \p_i (\r \xi_i)$. To obtain $\delta \O^2$ up to $O(R^{-3})
\times \O_{{\rm N}}^2$, we only need the following\footnote{In
  ref.\cite{PNeom} where Chandrasekhar calculated the PN configuration
  of the Jacobi ellipsoid, he introduced other type of displacements,
  such as $\xi^k (k=1 - 5)$. Coefficients of these displacements are,
  however, of $O(R^{-3})$ in this paper because the
  ellipsoidal displacements of a star are generated by the small spin of 
  each star and by the tidal
  force of the companion star both of which deform the stars only by
  $O(R^{-3})$. Then, the displacements will contribute
  to the orbital angular velocity from a higher order of $R^{-1}$ as
  $\O_{{\rm N}}^2 \times O(R^{-5})$. This is the reason why we do not
  need $\gamma_0$, $\gamma_i$ and $\gamma_{ij}$. }:
\beqa
  \xi_i^{(0)} &=& \left( \frac{1}{2}, 0, 0 \right), \label{Ld1}\\
  \xi_i^{(21)} &=& \left( \frac{1}{2} x_1^2, 0, -x_1 x_3 \right),
  \label{Ld2} \\
  \xi_i^{(22)} &=& (0, x_1 x_2, -x_1 x_3). \label{Ld3}
\eeqa
Thus, we set
\beqa
  \xi_i = S_0 \xi_i^{(0)} +S_{21} \xi_i^{(21)} +S_{22} \xi_i^{(22)},
\eeqa
where $S_0$, $S_{21}$ and $S_{22}$ are constants. 
In this case, we can get $\delta U$ of the contribution from star 1 as
\beqa
  \delta U^{1 \ra 1} &=& \pi \r A_1 S_0 x_1 -\left( \frac{1}{2} \frac{\p
      D_{11}}{\p x_1} - \frac{\p D_{13}}{\p x_3} \right) S_{21} -\left
    ( \frac{\p D_{12}}{\p x_2} -\frac{\p D_{13}}{\p x_3} \right) S_{22},
\eeqa
where
\beqa
  D_{1i} &=& \pi \r a_1^2 a_i^2 \left( A_{1i} -\sum_l A_{1il} x_l^2
  \right) x_1 x_i~~~(i \ne 1).
\eeqa
The contribution from star 2 is defined as
\beqa
  \delta U^{2 \ra 1} =-\sum_i \frac{\p}{\p x_i} \int_2 d^3 x' \frac{\r 
    \xi'_i}{|{\bf x}-{\bf x}'|}.
\eeqa
Here, the integral is performed in star 2, and $\xi'_i$ is defined
for star 2. Note that $|\xi'_i|$ is the same as $|\xi_i|$, but
$\xi'_i$ has opposite sign to $\xi_i$ if components of $\xi_i$ are the
even function of $x_i$. Using this definition, we have
\beqa
  \delta U^{2 \ra 1} &=&\frac{1}{2} S_0 \frac{\p U^{2 \ra 1}}{\p x_1}
  +\left( \frac{1}{2} \frac{\p D_{11}^{2 \ra 1}}{\p x_1} -\frac{\p
      D_{13}^{2 \ra 1}}{\p x_3} \right) S_{21} +\left( \frac{\p
      D_{12}^{2 \ra 1}}{\p x_2} -\frac{\p D_{13}^{2 \ra 1}}{\p x_3}
  \right) S_{22},\label{delueq}
\eeqa
where
\beqa
  D_{1i}^{2 \ra 1} &=& \frac{3I_{11ii}}{R^3} \left( \frac{x_i}{R}
    -\frac{4x_1x_i}{R^2} \right)~~~(i \ne 1).
\eeqa

In order to get the orbital angular velocity, we also have to calculate
the second and the last terms of the right-hand side of Eq.(\ref{PNftve})
which are occurred by the displacement of the fluid element by the PN
correction.  The second term is
\beqa
  \frac{M}{2} \O_{{\rm N}}^2 \left( S_0 + S_{21} \frac{a_1^2}{5} \right),
\eeqa
and the last one can be evaluated as follows: 
First, the contribution from star 1 is
zero because
\beqa
  \delta \int d^3 x \r \frac{\p U^{1 \ra 1}}{\p x_1} &=& -\delta \int d^3 x
  \r ({\bf x}) \int d^3 x' \r ({\bf x}') \frac{x_1-x_1'}{|{\bf x}-{\bf
      x}'|^3} \nonumber \\
  &=& -\int d^3 x \r ({\bf x}) \delta \int d^3 x' \r ({\bf x}')
  \frac{x_1-x_1'}{|{\bf x}-{\bf x}'|^3}
  +\int d^3 x' \r ({\bf x}')
  \delta \int d^3 x \r ({\bf x}) \frac{x_1'-x_1}{|{\bf x}-{\bf x}'|^3}
  \nonumber \\
  &=& 0. 
\eeqa
The contribution from star 2 becomes
\beqa
  \delta \int d^3 x \r \frac{\p U^{2 \ra 1}}{\p x_1} &=&
  \int d^3 x \r \sum_{i=1}^3 \xi_i \frac{\p^2 U^{2 \ra 1}}{\p x_1 \p
    x_i} + \int d^3 x \r \frac{\p \delta U^{2 \ra 1}}{\p x_1}.
  \label{delta_U^out}
\eeqa
The first term on the right hand side of Eq.(\ref{delta_U^out}) denotes
the force which the displaced element of star 1 receives from the
non-displaced potential of star 2. On the other hand, the second term
denotes the force which the non-displaced element of star 1 receives
from the displaced potential of star 2.
For the Lagrangian displacement given in Eqs.(\ref{Ld1})$-$(\ref{Ld3}), 
the first and the second terms of Eq.(\ref{delta_U^out}) are equal, and 
using Eq.(\ref{delueq}), we can calculate Eq.(\ref{delta_U^out}) as 
\beqa
  \delta \int d^3 x \r \frac{\p U^{2 \ra 1}}{\p x_1} &=&
  2 \int d^3 x \r \frac{\p \delta U^{2 \ra 1}}{\p x_1}, \\
  &=& S_0 \frac{2M}{R^3} \left( M +\frac{18 \bI_{11}}{R^2} \right)
  +S_{21} \frac{2M}{R^3} \left( I_{11} +\frac{9 \bI_{1111}}{R^2}
    +\frac{9 \bI_{11} I_{11}}{M R^2} +\frac{12 I_{1133}}{R^2} \right)
  \nonumber \\
  & &+S_{22} \frac{24M}{R^5} (I_{1133}-I_{1122}).
\eeqa

At the deformed surface of the Darwin ellipsoid, the pressure must
vanish. This condition becomes
\beqa
  \left( \frac{P}{\r} \right)_S &=& -2 \frac{P_0}{\r} \sum_l \frac{\xi_l 
    x_l}{a_l^2} +\delta U +\frac{1}{2} \varpi^2 \delta \O^2 \nonumber \\
  & &+ \gamma_0 +\sum_l \gamma_l x_l^2 +\sum_{l \ge m} \gamma_{lm} x_l^2
  x_m^2 +
  x_1 \left( \beta_0 +\sum_l \beta_l x_l^2 +\sum_{l \ge m} \beta_{lm}
    x_l^2 x_m^2 \right) +{\rm const.} \nonumber \\
  &=& 0, \label{sur_press}
\eeqa
where we use the equation of the boundary surface at the deformed Darwin
ellipsoid
\beqa
  S= \sum_l \frac{x_l^2}{a_l^2} -1 -2 \sum_l \frac{\xi_l x_l}{a_l^2} =0.
\eeqa
Eq.(\ref{sur_press}) must be satisfied at the original surface of the
Darwin ellipsoid, $\sum_l x_l^2/ a_l^2 =1$. Coefficients, $S_0$,
$S_{21}$ and $S_{22}$, are determined from the fact that the terms of
the odd power of $x_l$ must vanish. The coefficients of $x_1$, $x_1^3$,
$x_1 x_2^2$ and $x_1 x_3^2$ in Eq.(\ref{sur_press})
($Q_1$, $Q_{111}$, $Q_{122}$, and $Q_{133}$) are as follows:
\beqa
  Q_1 &=& \left( -\frac{P_0}{\r a_1^2} +\pi \r A_1 +\frac{\O_{\rm
        N}^2}{2} \right) S_0 +\frac{R}{2} \delta \O^2 +\beta_0 \nonumber
  \\
  & &+ S_{21} \pi \r a_1^2 \left( -a_1^2 A_{11} +\frac{1}{2} B_{11}
  +a_3^2 A_{13} +O(R^{-3}) \right) -S_{22} \pi \r a_1^2 \left(
   a_2^2 A_{12} -a_3^2 A_{13} \right), \label{Q_1} \\
  Q_{111} &=& \beta_1 +S_{21} \pi \r a_1^2 \left(
  -\frac{P_0}{\pi \r^2 a_1^4} +2a_1^2 A_{111} -a_3^2 A_{113}
  -\frac{1}{2} B_{111}  \right) \nonumber \\
  & &+S_{22} \pi \r a_1^2 \left( a_2^2 A_{112} -a_3^2 A_{113} \right),
  \\
  Q_{122} &=& \beta_2 +S_{21} \pi \r a_1^2 \left(
  a_1^2 A_{112} -\frac{1}{2} B_{112} -a_3^2 A_{123} \right) \nonumber \\
  & &+S_{22} \pi \r a_1^2 \left( -\frac{2P_0}{\pi \r^2 a_1^2 a_2^2}
  +3a_2^2 A_{122} -a_3^2 A_{123} \right), \\
  Q_{133} &=& \beta_3 +S_{21} \pi \r a_1^2 \left(
  \frac{2P_0}{\pi \r^2 a_1^2 a_3^2} +a_1^2 A_{113} -\frac{1}{2} B_{113} 
  -3a_3^2 A_{133} \right) \nonumber \\
  & &+S_{22} \pi \r a_1^2 \left( \frac{2P_0}{\pi \r^2 a_1^2 a_3^2}
  +a_2^2 A_{123} -3a_3^2 A_{133} \right).
\eeqa

The conditions to determine $S_0$, $S_{21}$ and $S_{22}$ are
\beqa
  Q_1 +Q_{133} a_3^2 &=& 0, \label{Q1} \\
  Q_{111} a_1^2 -Q_{133} a_3^2 &=& 0, \label{Q2} \\
  Q_{122} a_2^2 -Q_{133} a_3^2 &=& 0. \label{Q3}
\eeqa
It is found that the
coefficient of $S_0$ in $Q_1$ vanishes when we substitute
Eq.(\ref{PNftve}) to Eq.(\ref{Q1}) and remove $\delta \O^2$.
Thus, $S_0$ is
indeterminant, and the first equation becomes trivial one. This is also 
verified by calculating $3Q_{111} a_1^2 +Q_{122}a_2^2 +Q_{133} a_3^2$,
which coincides with $- 5 Q_1$. This fact seems to be very natural because
$S_0$ is a simple transformation of the center of mass. As a result, $S_{21}$
and $S_{22}$ are determined by the second and third equations. From 
the combination $2 \times (\ref{Q2}) - (\ref{Q3})$, i.e., 
$2Q_{111} a_1^2 -Q_{122} a_2^2 -Q_{133} a_3^2 =0$, 
we obtain
\beqa
  0 &=& 2\b_1 a_1^2 -\b_2 a_2^2 -\b_3 a_3^2 \nonumber \\
  & &+S_{21} \pi \r a_1^2 \Biggl[
    -\frac{4P_0}{\pi \r^2 a_1^2} +4a_1^4 A_{111} -a_1^2 a_2^2 A_{112}
    -3 a_1^2 a_3^2 A_{113} +a_2^2 a_3^2 A_{123} \nonumber \\
  & &\hspace{60pt} +3a_3^4 A_{133} -a_1^2
    B_{111} +\frac{1}{2}(a_2^2 B_{112} +a_3^2 B_{113}) \Biggr]
    \nonumber \\
  & &+S_{22} \pi \r a_1^2 \left[ 2a_1^2 (a_2^2 A_{112} -a_3^2 A_{113})
    -3a_2^4 A_{122} +3a_3^4 A_{133} \right],
\eeqa
where we omit higher order terms of $R^{-1}$. $2\b_1 a_1^2 -\b_2 a_2^2
-\b_3 a_3^2$ is calculated as
\beqa
  & &\frac{M \pi \r}{R^2} \left[ 2 a_1^2 A_1 -\frac{5}{2} a_2^2 A_2
    +\frac{1}{2} a_3^2 A_3 +\frac{5}{2} a_1^2 (a_2^2 A_{12} -a_3^2
    A_{13}) \right] +O(R^{-4}) \nonumber \\
  & &=O(R^{-4}),
\eeqa
where we use the relations
\beqa
  a_2^2 A_2 &=& a_1^2 A_1 +O(R^{-3}), \\
  a_3^2 A_3 &=& a_1^2 A_1 +O(R^{-3}), \\
  a_1^2 a_2^2 A_{12} &=& a_3^2 A_3 +O(R^{-3}), \\
  a_1^2 a_3^2 A_{13} &=& a_2^2 A_2 +O(R^{-3}).
\eeqa
The coefficient of $S_{22}$ is also of $O(R^{-3})$ because $a_2^2 A_{112}
-a_3^2 A_{113} = O(R^{-3})$ and $a_2^4 A_{122} -a_3^4 A_{133}
=O(R^{-3})$, but that of $S_{21}$ is $O(1)$.

On the other hand, from Eq.(\ref{Q3}), we have
\beqa
  0 &=& \beta_2 a_2^2 -\beta_3 a_3^2 \nonumber \\
  & &+S_{21} \pi \r a_1^2 \Biggl[ -{2 P_0
    \over \pi \r^2 a_1^2} +a_1^2 (a_2^2 A_{112} -a_3^2 A_{113}) -{1
    \over 2} (a_2^2 B_{112} -a_3^2 B_{113}) -a_2^2 a_3^2 A_{123} +3
  a_3^4 A_{133} \Biggr] \nonumber \\
  & &+S_{22} \pi \r a_1^2 \Biggl[ -{4P_0 \over \pi
    \r^2 a_1^2} +3 a_2^4 A_{122} -2 a_2^2 a_3^2 A_{123} +3 a_3^4 A_{133} 
  \Biggr].
\eeqa
The coefficients of $S_{21}$ and $S_{22}$ are of $O(1)$ in this equation,
and we can also show $\b_1 a_1^2 -\b_2 a_2^2 =O(R^{-4})$.
Thus, we can conclude that both $S_{21}$ and
$S_{22}$ are of $O(R^{-4})$.
Accordingly, we do not have to consider these Lagrangian
displacements in the following.

\subsection{The Post-Newtonian Orbital Angular Velocity}

Since $S_0$ is arbitrary, we set $S_0 =0$.
Hence, Eq.(\ref{PNftve}) is calculated as
\beqa
  -\frac{R}{2} M \delta \O^2 &=& M\beta_0 +3\beta_1 I_{11} +\beta_2
  I_{22} + \beta_3 I_{33} +5\beta_{11} I_{1111} +3\beta_{12} I_{1122}
  \nonumber \\
  & &+3\beta_{13} I_{1133} +\beta_{22} I_{2222} +\beta_{33} I_{3333}
  +\beta_{23} I_{2233}  \\
  &=& -\frac{M^2}{R^2} 2 \pi \r A_0 + \frac{9 M^3}{4 R^3} 
  -\frac{84}{5R^4} \pi \r \bI_{11} A_0 
      +\frac{M^3}{10R^5} (28a_1^2 -14a_2^2 -9a_3^2),
\eeqa
where we make use of the relations $A_0 =\sum_l A_l a_l^2$,
Eqs.(\ref{P_1})$-$(\ref{P_3}), equations for
$I_{iijj}$\footnote{$I_{iiii} =3M a_i^4/35$,~ $I_{iijj} =M a_i^2
  a_j^2/35~~(i \ne j)$}
and reduction formulas for $A_{ijk \cdots}$ 
and $B_{ijk \cdots}$ shown in the section 21 of Chandrasekhar's
textbook\cite{Ch69}.
Then, $\O^2$ is written as 
\beqa
  \O^2 &=& \frac{2M}{R^3} \left[ 1+ \frac{1}{c^2} \left\{ 2\pi \r A_0
   -\frac{9M}{4R} -\frac{M}{10R^3} (28a_1^2 -14a_2^2 -9a_3^2)
   +O(R^{-4}) \right\} \right] \nonumber \\
   & &+ \frac{18 \bI_{11}}{R^5} \left( 1+\frac{28}{15 c^2} \pi \r A_0
     +O(R^{-2}) \right).  \label{PNoav1}
\eeqa

Before proceeding further, we comment on the mass of the system and 
definition of the center of mass for each star. 
This is because there are several definitions of them in the PN 
approximation, and we should clarify the difference between the 
similar ones. 

First, we consider the conserved mass which is defined as
\beqa
  M_{\ast} &=& \int d^3 x \r \left[ 1+\frac{1}{c^2} \left( \frac{v^2}{2} 
      +3U \right) \right] \nonumber \\
  &=& M \left[ 1+\frac{1}{c^2} \left( \frac{13M}{4R} +\frac{12 \pi \r
        A_0}{5} +\frac{M}{20R^3} (34a_1^2 -11a_2^2 -15a_3^2) +O(R^{-5})
    \right) \right].\label{conmas}
\eeqa
Using $M_{\ast}$, $\O^2$ becomes
\beqa
  \O^2 &=& \frac{2M_{\ast}}{R^3} \left[ 1+
    \frac{1}{c^2} \left\{ -\frac{2\pi \r A_0}{5}
      -\frac{11M_{\ast}}{2R} 
  -\frac{M_{\ast}}{20R^3} (90a_1^2 -39a_2^2 -33a_3^2) \right\}
  +O(R^{-4}) \right] \nonumber \\
  & &+ \frac{18(\bI_{11})_{\ast}}{R^5} \left[ 1+\frac{1}{c^2} \left\{
      -\frac{8}{15} \pi \r A_0 -\frac{13 M_{\ast}}{4R} +O(R^{-2})
      \right\} \right],
\eeqa
where $(\bI_{11})_{\ast}=M_*\bI_{11}/M$. 
Thus, $\O^2$ looks as if it depends on the internal structure of the
star even for the limit $a_i/R \ra 0$. Since we believe that in the
equation of motion(EOM) for the point particle, the quantities
depending on the internal structure does not appear, $M_{\ast}$ is not
desirable to describe the EOM for the point particle. Instead, in the case
when the EOM is derived, one usually adopts the PPN mass\cite{Will}
defined as
\beqa
  M_{{\rm PPN}} &=& \int d^3 x \r \left[ 1+ \frac{1}{c^2} \left
      ( \frac{v^2}{2} +3U -\frac{1}{2} U_{{\rm self}}
    +\frac{v_{{\rm self}}^2}{2} \right) \right] \nonumber \\
  &=& M \left[ 1+\frac{1}{c^2} \left( \frac{13M}{4R} +2\pi \r A_0
      +\frac{M}{20R^3} (38a_1^2 -7a_2^2 -15a_3^2) +O(R^{-5}) \right) \right],
\eeqa
where $U_{\rm self}$ and $v_{\rm self}$ are the 
self gravity part of the Newtonian potential and the spin velocity of 
each star. 
When we rewrite Eq.(\ref{PNoav1}) using the PPN mass, the orbital
angular velocity does not depend on the internal structure of the star
and agrees with that for the point particle\cite{BDIWW} in the limit
$a_i/R \ra 0$ as
\beqa
  \O^2 &=& \frac{2M_{{\rm PPN}}}{R^3} \left[ 1
    +\frac{1}{c^2} \left\{ -\frac{11M_{{\rm PPN}}}{2R}
   -\frac{M_{{\rm PPN}}}{20 R^3}
   (94 a_1^2 -35a_2^2 -33a_3^2)
 +O(R^{-4}) \right\} \right] \nonumber \\
  & &+\frac{18(\bI_{11})_{{\rm PPN}}}{R^5} \left[ 1+\frac{1}{c^2}
    \left\{ -\frac{2}{15} \pi \r A_0 -\frac{13M_{{\rm PPN}}}{4R}
      +O(R^{-2}) \right\} \right], \label{PNoav3}
\eeqa
where $(\bI_{11})_{\rm PPN}=M_{\rm PPN}\bI_{11}/M$.
Thus, when we compare the present results with the point 
particle calculations, we should use the PPN mass. 
In the present case, however,
$M_{{\rm PPN}}$ is not a conserved quantity although $M_{\ast}$ is. When 
we consider a sequence of the equilibrium configuration as an
evolutionary sequence, we should fix $M_{\ast}$.

Next, we consider the definition of the center of mass for each star. 
In the PPN formalism, it is defined as\cite{Will}
\beqa
  x^i_{{\rm PPN}} =\frac{1}{M_{{\rm PPN}}}\int d^3 x \r x^i \left[ 1+
    \frac{1}{c^2} \left(
    \frac{v^2}{2} +3U -\frac{1}{2} U_{{\rm self}}
    +\frac{v_{{\rm self}}^2}{2} \right) \right],
\eeqa
and the $x_1$ coordinate of the center of mass for star 1 deviates from
0 to
\beqa
  \frac{1}{c^2} \left( -\frac{2M a_1^2}{5R^2} +O(R^{-4}) \right).
\eeqa
Thus, in the PPN formalism, the following orbital separation should be used;
\beqa
  R_{{\rm PPN}} =R \left[ 1+ \frac{1}{c^2} \left\{ -\frac{4M
        a_1^2}{5R^3} +O(R^{-5}) \right\} \right].
\eeqa

It is worth noting that 
when we define the center of mass by the conserved mass as
\beqa
  x^i_{\ast} =\frac{1}{M_{\ast}} \int d^3 x \r_{\ast} x^i,
\eeqa
the result is the same up to $O(R^{-4})$. 
Thus, in this paper, we do not have to distinguish $R_*$ from $R_{\rm PPN}$. 
Even in the general cases,  the difference between 
$R_*$ and $R_{\rm PPN}$ is expected to be small. 

Using $R_*$ and/or $R_{\rm PPN}$, $\O^2$ is rewritten as 
\beqa
  \O^2 &=& \frac{2M_{\ast}}{R_{\ast}^3} \left[ 1+
    \frac{1}{c^2} \left\{ -\frac{2\pi \r A_0}{5}
      -\frac{11M_{\ast}}{2R_{\ast}} 
  -\frac{M_{\ast}}{20R_{\ast}^3} (138a_1^2 -39a_2^2 -33a_3^2) \right\}
  +O(R_{\ast}^{-4}) \right] \nonumber \\
  & &+ \frac{18(\bI_{11})_{\ast}}{R_{\ast}^5} \left[ 1+\frac{1}{c^2}
    \left\{ -\frac{8}{15} \pi \r A_0 -\frac{13 M_{\ast}}{4R_{\ast}}
      +O(R_{\ast}^{-2}) \right\} \right],
\eeqa
or
\beqa
  \O^2 &=& \frac{2M_{{\rm PPN}}}{R_{{\rm PPN}}^3} \left[ 1
    +\frac{1}{c^2} \left\{ -\frac{11M_{{\rm PPN}}}{2R_{{\rm PPN}}}
   -\frac{M_{{\rm PPN}}}{20 R_{{\rm PPN}}^3}
   (142 a_1^2 -35a_2^2 -33a_3^2)
 +O(R_{{\rm PPN}}^{-4}) \right\} \right] \nonumber \\
  & &+\frac{18(\bI_{11})_{{\rm PPN}}}{R_{{\rm PPN}}^5} \left[
   1+\frac{1}{c^2} \left\{ -\frac{2}{15} \pi \r A_0
   -\frac{13M_{{\rm PPN}}}{4R_{{\rm PPN}}} 
      +O(R_{{\rm PPN}}^{-2}) \right\} \right].\label{omepnp}
\eeqa
Here, we should note that the effect of the spin-orbit coupling terms 
will appear in $\O^2$ from $O(R^{-6}_{\rm PPN})$\cite{kidder}. 
According to the PN study, it becomes 
\beq
\O^2 = {2M_{\rm PPN} \over R_{\rm PPN}^3 }
\left[ 1-{2 M_{\rm PPN} \over  c^2R_{\rm PPN}^3}(a_1^2+a_2^2)\right],
\label{ooooo}
\eeq
where we omit other terms which do not concern in this discussion. 
Eq.(\ref{ooooo}) shows that the terms of $O(R^{-6}_{\rm PPN})$ in 
Eq.(\ref{omepnp}) cannot be explained only by the spin-orbit 
coupling term. This means that there is a new effect, say the PN 
quadrupole one, in Eq.(\ref{omepnp}). 

\section{The energy and the angular momentum}

\subsection{The Total Energy}

The PN total energy is calculated from $E=E_{{\rm N}}^{{\rm
    def}}+E_{{\rm PN}}^{{\rm def}}/c^2$, where\cite{chand}\cite{shibapn}
\beqa
  E_{{\rm N}}^{{\rm def}} &=& \int \r \left( \frac{1}{2} v^2
    -\frac{1}{2} U \right) d^3 x, \\
  E_{{\rm PN}}^{{\rm def}} &=& \int \r \left( \frac{5}{8} v^4
    +\frac{5}{2} v^2 U +\frac{P}{\r} v^2 -\frac{5}{2} U^2 +\frac{1}{2}
    \hat{\beta}_{\varphi} \O^2 \right) d^3 x.
\eeqa
For the PN Darwin problem, they become
\beqa
  E_{{\rm N}}^{{\rm def}} &=& M \left[ -\frac{4\pi \r A_0}{5}
    +\frac{\O^2}{5} (a_1^2 +a_2^2) +\frac{R^2 \O^2}{4} -\frac{M}{R}
    -\frac{3 \bI_{11}}{R^3} +O(R^{-5}) \right], \label{neneeq} \\
  E_{{\rm PN}}^{{\rm def}} &=& 2M \Biggl[ -{1 \over 7} (\pi \r)^2
    (11A_0^2 +a_1^4 A_1^2 +a_2^4 A_2^2 +a_3^4 A_3^2) -{4 M \pi \r A_0
    \over R} \nonumber \\
  & &\hspace{30pt} -{M \pi \r \over 7R^3} \left\{ {87 \bI_{11} \over M}
    A_0 -(2 a_1^4 A_1 -a_2^4 A_2 -a_3^4 A_3) \right\} \nonumber \\
  & &\hspace{30pt} -{5 \over 2} \left\{
    {M^2 \over R^2} +{M^2 \over 5R^4} (5a_1^2 -2a_2^2 -2a_3^2)
    \right\} \nonumber \\
  & &\hspace{30pt} +\O^2 \left\{ {17 \over 16} MR +{M \over 80R} (6
    a_1^2 +21 a_2^2 -17 a_3^2) \right\} \nonumber \\
  & &\hspace{30pt} +\O^2 \pi \r \Biggl\{ {R^2 \over 20} (3A_0
    -a_2^2 A_2) +{3 \over 7} A_0 (a_1^2 +a_2^2) -{12 \over 35} (a_1^4
    A_1 +a_2^4 A_2) \nonumber \\
  & &\hspace{80pt} -{1 \over 35} a_1^2 a_2^2 (A_1 +A_2) \Biggr\}
   \nonumber \\
  & &\hspace{30pt} +{P_0
    \over \r} \O^2 \left\{ {R^2 \over 10} +{2 \over 35} (a_1^2 +a_2^2)
    \right\} +{5 \over 8} \O^4 \left\{ {R^4 \over 16} +{R^2 \over 10}
    (3a_1^2 +a_2^2) \right\} +O(R^{-5}) \Biggr], \label{eneeq1} \\
  &=& 2M \Biggl[ -\frac{34}{21} (\pi \r A_0)^2
    -\frac{11M \pi \r A_0}{3R} -\frac{7M^2}{32R^2} +\frac{M \pi \r
    A_0}{R^3} \left\{ \frac{68}{105} (a_1^2+a_2^2) -\frac{61}{7}
    \frac{\bI_{11}}{M} \right\} \nonumber \\
  & &\hspace{50pt} +\frac{M^2}{240R^4} (302a_1^2 +59a_2^2
    -209a_3^2) +O(R^{-5}) \Biggr], \label{eneeq}
\eeqa
where we use the equilibrium equations (\ref{P_0})$-$(\ref{P_3}) and
    $\O^2$ to reduce Eq.(\ref{eneeq1}) to Eq.(\ref{eneeq}).

If we substitute $\O^2$ into the above formulas, $E$
may be rewritten as $E_{{\rm N}} + E_{{\rm PN}}/c^2$ where
\beqa
  E_{{\rm N}} &=& M \left[ -\frac{4\pi \r A_0}{5}
    -\frac{M}{2R}
    +\frac{3 \bI_{11}}{2R^3}
    +\frac{\O_{{\rm N}}^2}{5} (a_1^2 +a_2^2)
    +O(R^{-5}) \right], \\
  E_{{\rm PN}} &=& 2M \left[ -\frac{34}{21} (\pi \r A_0)^2
    -\frac{19M \pi \r A_0}{6R} -\frac{25M^2}{32R^2} +\frac{M \pi \r
    A_0}{R^3} \left\{ \frac{22}{21} (a_1^2+a_2^2) -\frac{158}{35}
    \frac{\bI_{11}}{M} \right\} \right. \nonumber \\
  & &\hspace{50pt} \left. +\frac{M^2}{240R^4} (26a_1^2 +35a_2^2
    -155a_3^2) +O(R^{-5}) \right].
\eeqa

\subsection{The Total Angular Momentum}

We can calculate the PN total angular momentum from 
$J=J_{{\rm N}}^{{\rm def}} +J_{{\rm PN}}^{{\rm def}}/c^2$, 
where\cite{chand}\cite{shibapn}
\beqa
  J_{{\rm N}}^{{\rm def}} &=& \int \r v_{\varphi} d^3 x, \\
  J_{{\rm PN}}^{{\rm def}} &=& \int \r \left[ v_{\varphi} \left( v^2 +6U 
    +\frac{P}{\r} \right) + \hat{\b_{\varphi}} \O \right] d^3 x, 
\eeqa
and $v_{\varphi} = \O \varpi^2$.
For the PN Darwin problem, they become
\beqa
  J_{{\rm N}}^{{\rm def}} &=& 2M \O \left( \frac{R^2}{4} + \frac{a_1^2
    +a_2^2}{5} \right), \label{nangeq} \\
  J_{{\rm PN}}^{{\rm def}} &=& M \O \Biggl[ R^2 \pi \r A_0 +{P_0 \over
    \r} \left\{ {R^2 \over 5} +{4 \over 35} (a_1^2 +a_2^2) \right\}
  -{R^2 \over 5} \pi \r a_2^2 A_2 \nonumber \\
  & &\hspace{30pt} +{4 \pi \r \over 35} \Bigl\{ 18 A_0
  (a_1^2 +a_2^2) -13 (a_1^4 A_1 +a_2^4 A_2) -a_1^2 a_2^2 (A_1+A_2)
  \Bigr\} \nonumber \\
  & &\hspace{30pt} +\O^2 \left\{ {R^4 \over 8} +{R^2 \over 5} (3a_1^2 +a_2^2)
  \right\} +{19 M R \over 4} +{M \over 20 R} (10 a_1^2 +27 a_2^2 -19
  a_3^2) \Biggr], \label{angeq1} \\
  &=& M \O_{{\rm N}} \Biggl[ R^2 \pi \r A_0 +5 R 
    M + \frac{164}{105} \pi \r A_0 (a_1^2+a_2^2) \nonumber \\
  & &\hspace{50pt} +\frac{M}{10R} (20a_1^2 
    +15a_2^2 -11a_3^2) +O(R^{-2}) \Biggr], \label{angeq}
\eeqa
where we use the equilibrium equations (\ref{P_0})$-$(\ref{P_3}) and
    $\O^2$ to reduce Eq.(\ref{angeq1}) to Eq.(\ref{angeq}).

If we substitute $\O^2$ into the above formulas, $J$ may be rewritten as 
$J_{{\rm N}}+ J_{{\rm PN}}/c^2$ where
\beqa
  J_{{\rm N}} &=& 2M \O_{{\rm N}} \left( \frac{R^2}{4} + \frac{a_1^2
    +a_2^2}{5} \right), \\
  J_{{\rm PN}} &=& M \O_{{\rm N}} \left[ \frac{3}{2} R^2 \pi \r A_0
    +\frac{71}{16} R M +\frac{\pi \r A_0}{1050} (2018 a_1^2 +2081 a_2^2
    +21a_3^2) \right. \nonumber \\
  & &\hspace{50pt} \left.+\frac{M}{80 R} (122a_1^2 +85 a_2^2 -97 a_3^2)
    +O(R^{-2}) \right].
\eeqa

\section{Equilibrium sequence of the post-Newtonian Darwin ellipsoid}

In this section, we construct the equilibrium sequences of the PN Darwin
ellipsoids fixing $M_*$ and $\rho$ as follows;

\noindent
(1) using Eqs.(\ref{P_1})$-$(\ref{P_3}),
we numerically calculate the equilibrium sequences of the Newtonian order. 
Up to this stage, $\alpha_2$, $\alpha_3$, and $\tilde R=R/a_1$ 
are determined. 

\noindent
(2) $a_1$ is determined from the condition $M_*=$constant 
using Eq.(\ref{conmas}):
\beq
a_1^3={3 M_* \over 4\pi\rho\alpha_2\alpha_3}
\biggl[1-{\pi\rho \over c^2}\biggl({3M_* \over 4\pi\rho\alpha_2\alpha_3}
\biggr)^{2/3}\biggl({12 \tilde A_0 \over 5}
+{13 \alpha_2\alpha_3 \over 3\tilde R}
+{\alpha_2\alpha_3 \over 15 \tilde R^3}(34-11\alpha_2^2-15\alpha_3^2)\biggr)
\biggr].\label{aaeq}
\eeq

\noindent
(3) after 
substituting Eq.(\ref{aaeq}) into Eqs.(\ref{PNoav1}), (\ref{neneeq}),
(\ref{eneeq}), (\ref{nangeq}),
and (\ref{angeq}), we rewrite the PN expressions for the orbital
angular velocity, the energy and the angular momentum as
\beqa
  \tilde{\O}^2 &\equiv& \frac{\O^2}{\pi \r}
  = \tilde{\O}_{\rm N}^2 +\frac{M_{\ast}}{c^2 a_{\ast}}
  \tilde{\O}_{\rm PN}^2, \\
  \tilde{E} &\equiv& \frac{E}{(M_{\ast}^2/a_{\ast})}
  = \tilde{E}_{\rm N} +\frac{M_{\ast}}{c^2 a_{\ast}}
  \tilde{E}_{\rm PN},\\
  \tilde{J} &\equiv& \frac{J}{(M_{\ast}^3 a_{\ast})^{1/2}}
  = \tilde{J}_{\rm N} +\frac{M_{\ast}}{c^2 a_{\ast}}
  \tilde{J}_{\rm PN},
\eeqa
where
\beqa
  a_{\ast} &=&\left( \frac{3M_{\ast}}{4 \pi \r} \right)^{1/3},\\
  \tilde{\O}_{\rm N}^2 &=& \frac{4}{3} \a_2 \a_3 \left[
   \frac{2}{\tilde{R}^3} +\frac{6}{5\tilde{R}^5} (2-\a_2^2-\a_3^2)
 \right], \\
  \tilde{\O}_{\rm PN}^2 &=& \frac{4}{3} (\a_2 \a_3)^{4/3} \Biggl[
   \frac{3}{\tilde{R}^3} \frac{\tilde{A_0}}{\a_2 \a_3}
   -\frac{9}{2\tilde{R}^4} +\frac{42}{25\tilde{R}^5}
   \frac{\tilde{A_0}}{\a_2 \a_3} (2-\a_2^2-\a_3^2) \nonumber \\
  & &\hspace{70pt}-\frac{1}{5\tilde{R}^6}(28-14\a_2^2-9\a_3^2)
  \Biggr], \\
  \tilde{E}_{\rm N} &=& (\a_2\a_3)^{1/3} \Biggl[ -\frac{3}{5}
    \frac{\tilde{A_0}}{\a_2 \a_3}
  -\frac{1}{2 \tilde{R}} +\frac{1}{10 \tilde{R}^3}(2-\a_2^2-\a_3^2)
  \nonumber \\
  & &\hspace{60pt}+
  \frac{3\tilde \Omega_{\rm N}^2}{20\alpha_2\alpha_3} (1+\a_2^2) \Biggr], \\
  \tilde{E}_{\rm PN} &=& (\a_2\a_3)^{2/3} \Biggl[
  -\frac{3}{140} \left( \frac{\tilde{A_0}}{\a_2
      \a_3} \right)^2 +\frac{55}{48 \tilde{R}^2} +\frac{1}{700
    \tilde{R}^3} \frac{\tilde{A_0}}{\a_2 \a_3} (398+401\a_2^2+\a_3^2)
  \nonumber \\
  & &\hspace{60pt} -\frac{1}{120
      \tilde{R}^4}(194+215\a_2^2+165\a_3^2) \Biggr], \\
  \tilde{J}_{\rm N} &=& 2 (\a_2\a_3)^{-1/6} \left[
   \frac{3\tilde \Omega_{\rm N}^2}{4\alpha_2\alpha_3} 
 \right]^{1/2} \left( \frac{\tilde{R}^2}{4}+\frac{1+\a_2^2}{5} \right),
 \\
  \tilde{J}_{\rm PN} &=& (\a_2\a_3)^{1/6} \left[
   \frac{3\tilde \Omega_{\rm N}^2}{4\alpha_2\alpha_3}
 \right]^{1/2} 
  \Biggl[ -\frac{3}{8} \tilde{R}^2 \frac{\tilde{A_0}}{\a_2
     \a_3} +\frac{83}{48} \tilde{R} \nonumber \\
  & & +\frac{1}{1400}
   \frac{\tilde{A_0}}{\a_2 \a_3} (338+401\a_2^2+21\a_3^2)
 -\frac{1}{240\tilde{R}} (494+155\a_2^2+141\a_3^2) \Biggr].
\eeqa
 Then, 
using the numerical value of $\alpha_2$, $\alpha_3$, and $\tilde R$ 
determined at (1), we calculate the sequence of the angular velocity, 
the energy and the angular momentum as a function of the
orbital separation. 

We repeat this procedure changing the mean radius of each star $a_*$. 
Once a sequence is obtained, we 
search the minimum point of the energy. 
If we find it, we call it the ISCCO.

In figs. 2(a) and (b), we show $\tilde{E}$ and $\tilde{J}$ as functions of 
the normalized separation $R_{\ast}/a_{\ast}$, 
where $\tilde{E} =E/(M_{\ast}^2/a_{\ast})$ and
$\tilde{J}= J/(M_{\ast}^3 a_{\ast})^{1/2}$.
The figures show the important fact that the 
orbital separation at the ISCCO decreases approximately in proportion to 
$M_{\ast}/c^2 a_{\ast}$, i.e., the characteristic value of the 
compactness of each star. 

In fig. 3, we show the normalized orbital angular velocity 
$\tilde{\O}=\O/\sqrt{\pi \r}$ at the ISCCO as a function of
compactness. We show $\tilde{\O}$ at the energy minimum as well as 
that at the angular momentum minimum. The figure indicates that 
two minimums are almost coincident, but slightly different. 
The deviation between the location of the minimums of the energy and the 
angular momentum comes from the fact that we assume $a_{\ast}/R_{\ast}$
is a small parameter, and expand the energy to $O(R^{-4})$ and the
angular momentum to $O(R^{-1})$.
In any case, we may expect that the ISCCO locates near two minimums. 
Fig. 3 clearly shows that the orbital angular velocity at the ISCCO 
increases almost linearly with increase of the compactness. 
Since the PN approximation is valid only for small compactness, we can 
not do solid estimate of $\tilde \Omega$ at the ISCCO for a 
realistic NS of $M_{\ast}/c^2 a_{\ast} \sim 0.2$. 
Here, we dare to extrapolate this results to the 
realistic NS's. Then, we can 
find that the angular velocity at the ISCCO is 
about 10$\%$ larger than that in the Newtonian case. 

Before closing this section, we summarize the numerical results in Table I.
In Table I, $\dagger$ denotes the point of the ISCCO defined 
by the energy minimum. We note that in the Newtonian order,
the axial ratios $\alpha_2$ and $\alpha_3$
at the ISCCO are 0.8573 and 0.7995,
respectively\cite{LRS}. Thus, the PN effect increases $\alpha_2$ and 
$\alpha_3$ at the ISCCO; i.e., tidal deformation is weak compared with the 
Newtonian case. The reason for such a 
behavior seems to be that each star of binary is forced to be compact 
due to the PN gravity(see Eq.(\ref{aaeq})), and 
as a result, the tidal force is less effective 
than in the Newtonian case. 

\section{Summary}

In this paper, we have calculated the equilibrium sequences of the 
corotating BNS's of the incompressible fluid using the
first PN approximation of general relativity. The conclusions are as follows.

\noindent
(1) due to the PN effect, the orbital separation at the ISCCO(secular 
instability limit) decreases in proportion 
to the compactness of the star $M_*/c^2a_*$. 

\noindent
(2) the orbital angular velocity at the ISCCO increases in proportion 
to the compactness of the star $M_*/c^2a_*$. 

\noindent
(3) the reason for features (1) and (2) is that each star is forced to 
be compact due to the PN gravity and as a result, the tidal effect 
becomes less effective compared with the Newtonian binary. 

\noindent
These results agree with the recent numerical study by Shibata\cite{shibapn}. 

Since the analysis in this paper is done almost analytically, the 
equilibrium sequence obtained may be regarded as the exact solution 
of the Einstein equation for the limit of small compactness 
$M_*/c^2a_* \ll 1$ and $a_* \ll R$. 
Recently, two groups have been performed numerical 
computations for obtaining the equilibrium state of BNS's 
using the semi-relativistic approximation\cite{wilson}\cite{cornell}. 
The present solution is useful to check their numerical solutions as well 
as to understand the essence in their results. 

Finally, we comment on the possibility of extending this work to 
the compressible fluid case. In this paper, it is found that 
the deformation of the ellipsoidal figure due to 
the PN gravity is small effect to the 
angular velocity, so that the ellipsoidal model for the 
density configuration may be a good approximation. 
If we determine the density configuration as the ellipsoidal 
approximation as was done by Lai, Rasio and Shapiro\cite{LRS}, 
we only need to carry out the integrals which appear (a) in calculating 
the angular velocity by the first TV equation , and (b) in calculating 
the energy and the angular momentum, which can be done easily. 
We will publish results of such a calculation 
in a subsequent paper\cite{shibatani}. 

\acknowledgments 

We thank H. Asada and T. Tanaka for useful discussions and also K.T.
would like to thank professors 
H. Sato and T. Nakamura for helpful comments and 
continuous encouragement. This work was
in part supported by a Grant-in-Aid of Ministry of Education, Culture,
Science and Sports (Nos. 08237210 and 08NP0801).

\vspace{1.cm}

\begin{center}
  {\large TABLE CAPTIONS}
\end{center}

\vskip 5mm

Table.I. Equilibrium sequences of the PN Darwin ellipsoids
with $M_{\ast}/c^2a_{\ast}=0.01$, 0.03 and 0.05. $\dagger$ denotes the point
of the ISCCO.

\begin{table}
 \begin{center}
  \begin{tabular}{lcccccc}
    $R_{\ast}/a_{\ast}$&$R_{\ast}/(a_1)_{\ast}$&$a_2/a_1$&$a_3/a_1$&
    $\tilde{\Omega}$&$\tilde{J}$&$\tilde{E}$\\ \hline
   \multicolumn{7}{c}{$M_{\ast}/c^2a_{\ast}=0.01$} \\ \hline
   2.80&2.190&0.7283&0.6571&0.3645&1.574&-1.303 \\
   3.00&2.542&0.8135&0.7483&0.3208&1.527&-1.317 \\
   3.202$\dagger$&2.829&0.8593&0.8020&0.2877&1.518&-1.320 \\
   3.25&2.891&0.8673&0.8119&0.2809&1.518&-1.320 \\
   3.50&3.205&0.8999&0.8532&0.2497&1.528&-1.317 \\
   4.00&3.785&0.9372&0.9040&0.2033&1.571&-1.309 \\ \hline
   \multicolumn{7}{c}{$M_{\ast}/c^2a_{\ast}=0.03$} \\ \hline
   2.80&2.296&0.7776&0.7087&0.3488&1.534&-1.311 \\
   3.00&2.602&0.8388&0.7775&0.3101&1.510&-1.318 \\
   3.123$\dagger$&2.768&0.8633&0.8069&0.2906&1.507&-1.319 \\
   3.25&2.931&0.8828&0.8312&0.2729&1.509&-1.318 \\
   3.50&3.235&0.9106&0.8673&0.2433&1.521&-1.315 \\
   4.00&3.805&0.9433&0.9126&0.1987&1.565&-1.307 \\ \hline
   \multicolumn{7}{c}{$M_{\ast}/c^2a_{\ast}=0.05$} \\ \hline
   2.80&2.370&0.8121&0.7467&0.3353&1.508&-1.314 \\
   3.00&2.651&0.8596&0.8024&0.2999&1.496&-1.318 \\
   3.044$\dagger$&2.708&0.8674&0.8119&0.2932&1.495&-1.318 \\
   3.25&2.966&0.8962&0.8484&0.2649&1.500&-1.316 \\
   3.50&3.262&0.9201&0.8801&0.2368&1.514&-1.313 \\
   4.00&3.824&0.9488&0.9206&0.1939&1.558&-1.305
  \end{tabular}
 \end{center}
 \caption{}
 \label{table1}
\end{table}%

\newpage

\begin{center}
  {\large FIGURE CAPTIONS}
\end{center}

\vspace{0.5cm}

Fig.1. Sketch of the Darwin ellipsoid. The origin of the coordinate we
choose locates at the center of mass of star 1.

Fig.2(a). The total energy of the equilibrium sequence as a function of
$R_{\ast}/a_{\ast}$. Solid, dotted, dashed and long-dashed lines denote
$M_{\ast}/c^2a_{\ast}=0$(the Newtonian case), 
0.01, 0.03 and 0.05, respectively.

Fig.2(b). The total angular momentum of the equilibrium sequence as a
function of $R_{\ast}/a_{\ast}$. Solid, dotted, dashed and long-dashed
lines denote $M_{\ast}/c^2a_{\ast}=0$(the Newtonian case), 0.01, 0.03 and 0.05,
respectively.

Fig.3. The orbital angular velocity at the ISCCO as a function of the
compactness parameter $M_{\ast}/c^2a_{\ast}$. Filled circles and open
triangles denote $\O/\sqrt{\pi \r}$ for the minimum points of the total energy
and total angular momentum, respectively.

\newpage

\begin{figure}[ht]
  \vspace{1cm}
  \centerline{\epsfysize 10cm \epsfxsize 15cm \epsfbox{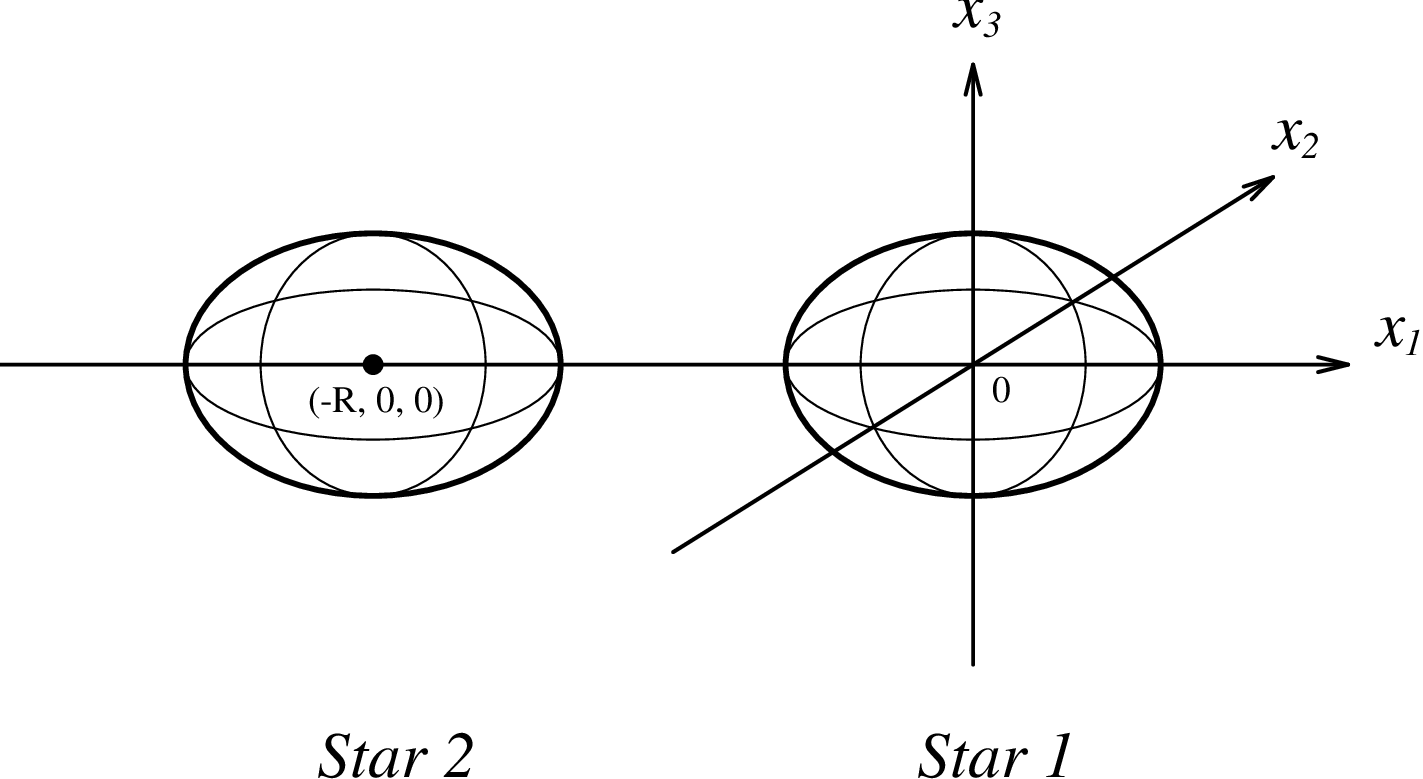}}
  \vspace{0.5cm}
  \label{darwin}
\end{figure}%

\vspace{3cm}

\hspace{250pt}
K. Taniguchi \& M. Shibata

\hspace{250pt}
Fig.1

\newpage

\begin{figure}[ht]
  \vspace{1cm}
  \centerline{\epsfysize 15cm \epsfxsize 15cm \epsfbox{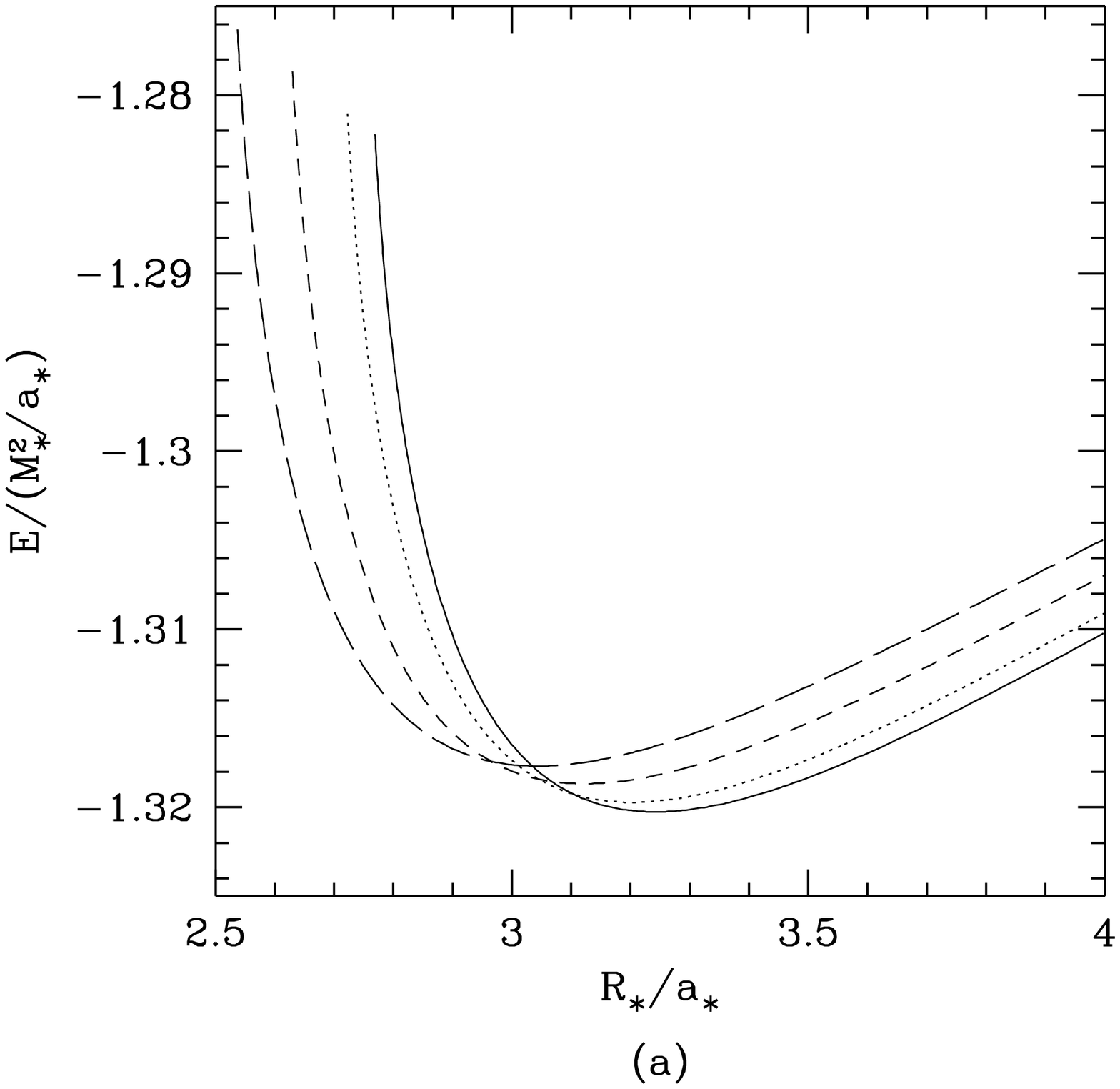}}
  \vspace{0.5cm}
  \label{energy}
\end{figure}%

\vspace{3cm}

\hspace{250pt}
K. Taniguchi \& M. Shibata

\hspace{250pt}
Fig.2(a)

\newpage

\begin{figure}[ht]
  \vspace{1cm}
  \centerline{\epsfysize 15cm \epsfxsize 15cm \epsfbox{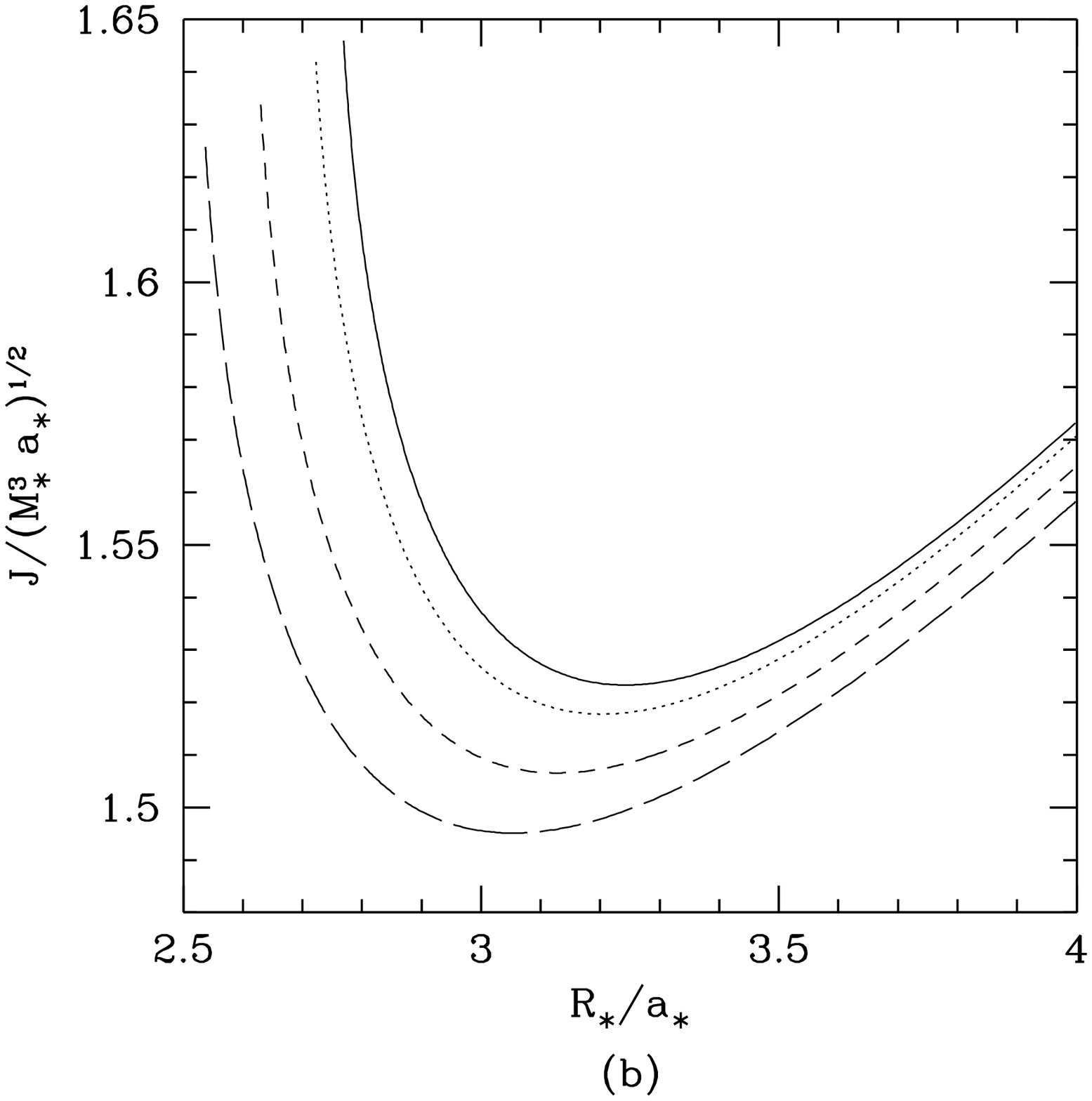}}
  \vspace{0.5cm}
  \label{angmom}
\end{figure}%

\vspace{3cm}

\hspace{250pt}
K. Taniguchi \& M. Shibata

\hspace{250pt}
Fig.2(b)

\newpage

\begin{figure}[ht]
  \vspace{1cm}
  \centerline{\epsfysize 15cm \epsfxsize 15cm \epsfbox{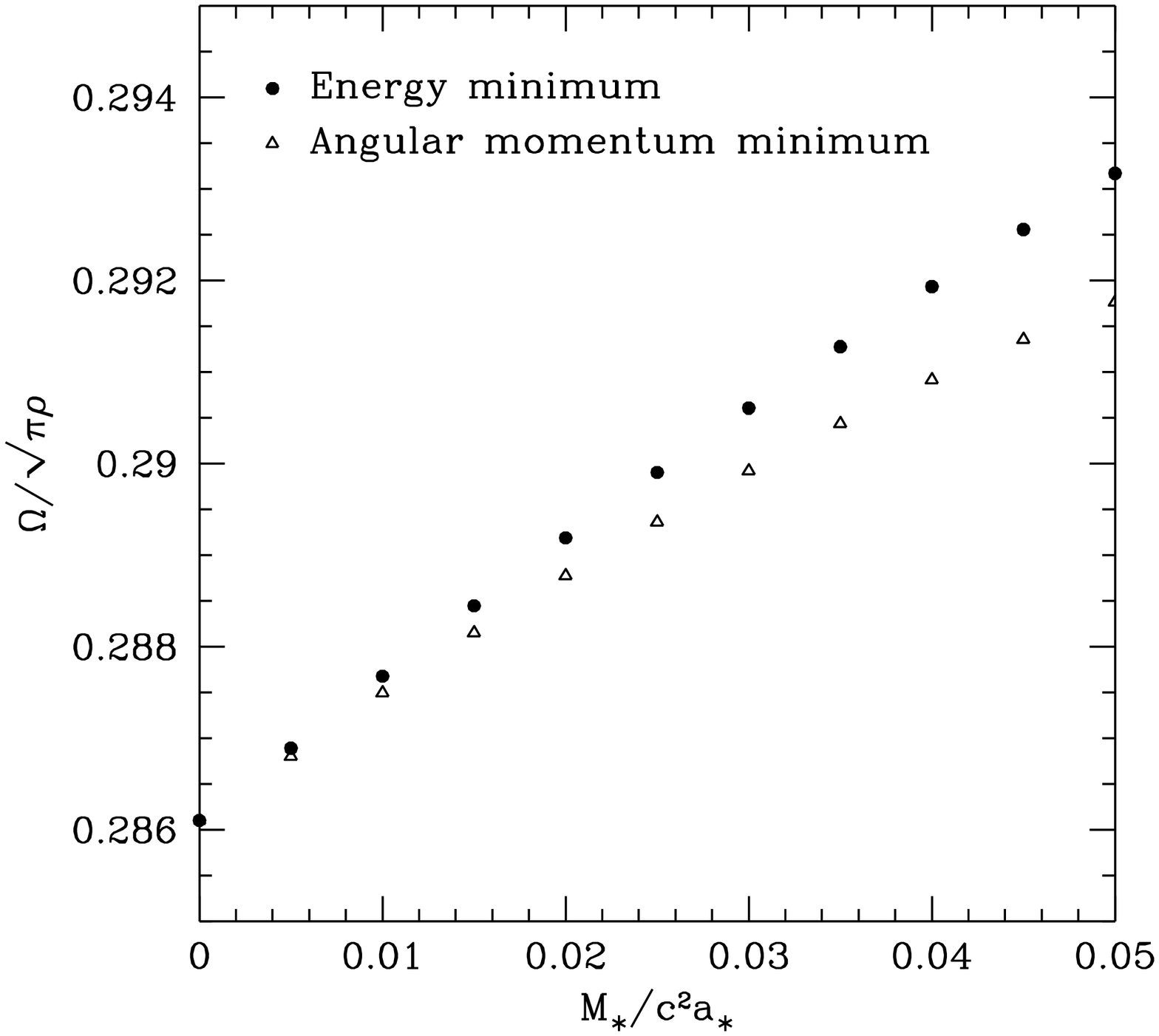}}
  \vspace{0.5cm}
  \label{omega}
\end{figure}%

\vspace{3cm}

\hspace{250pt}
K. Taniguchi \& M. Shibata

\hspace{250pt}
Fig.3

\end{document}